\documentclass{elsart}

\usepackage{graphicx}
\usepackage{amssymb}
\graphicspath{{.}{./EPS/}}

\begin{document}

\begin{frontmatter}

\title{Rashba spin splitting in quantum wires}

\author{M. Governale\corauthref{corinfo}}
\address{Scuola Normale Superiore, Piazza dei Cavalieri 7,
I-56126 Pisa, Italy}
\ead{governale@sns.it}
\corauth[corinfo]{Corresponding author. {\it Tel.:\/} 
+39 050 509258, {\it fax:}\/
+39 050 563513.}

\author{U. Z\"ulicke}
\address{Institute of Fundamental Sciences, Massey University,
Private Bag 11~222, Palmerston North, New Zealand}

\begin{abstract}
This article presents an overview of results pertaining to
electronic structure, transport properties, and interaction
effects in ballistic quantum wires with Rashba spin splitting.
Limits of weak and strong spin--orbit coupling are
distinguished, and spin properties of the electronic states
elucidated. The case of strong Rashba spin splitting where the
spin--precession length is comparable to the wire width turns
out to be particularly interesting. Hybridization of
spin--split quantum--wire subbands leads to an unusual spin
structure where the direction of motion for electrons can fix
their spin state. This peculiar property has important
ramifications for linear transport in the quantum wire, giving
rise to spin accumulation without magnetic fields or
ferromagnetic contacts. A description for interacting
Rashba--split quantum wires is developed, which is based on a
generalization of the Tomonaga--Luttinger model.      
\end{abstract}

\end{frontmatter}

\section{Introduction}

Spin--dependent transport in nanostructures has attracted a
lot of interest~\cite{sciencerev,lossbook} recently. A
familiar example are magnetoresistance effects in hybrid
systems~\cite{prinz:sc:90} consisting of magnetic and
nonmagnetics parts, which have important applications~\cite
{prinz:sc:98} in present data--storage technology. In such
magnetoelectronics devices, spin--dependent conductances arise
due to the interplay between ferromagnetic exchange--field
splitting and the Pauli principle. A finite spin polarization
of electric current {\em in the normal parts of a hybrid
system\/} is a prerequisite~\cite{fert:prb:93} for
magnetoresistance to occur. Strong efforts are currently
directed towards doing magnetoelectronics using the recently
discovered diluted magnetic semiconductor materials~\cite
{ohno:sci:98}.

Parallel to the pursuit of a semiconductor--magnetoelectronics
paradigm, recent studies have focused on finding out how the
quantum nature of spin can affect current flow, in particular,
in {\em nonmagnetic\/} semiconductor nanostructures. The
fundamentally relativistic coupling between charge carriers'
spin and orbital degrees of freedom turns out to give rise to
a host of interesting, and sometimes counterintuitive,
spin--dependent transport effects. Of special appeal is the
Rashba spin splitting~\cite{rashba,byra:jetplett:84} arising
in semiconductor heterostructures due to structural inversion
asymmetry~\cite{roess:prl:88,wink:prb:00}. The possibility to
tune its strength by external gate voltages was demonstrated
experimentally~\cite{nitta:prl:97,schaep:prb-rc:97} and forms
the basis for a spin--dependent field--effect--transistor
(spinFET) design~\cite{spinfet}. Early studies~\cite
{levitov:jetp:85,edel:ssc:90} discussed magnetoelectric 
effects in two--dimensional electron systems with Rashba spin
splitting. (See also recent related work~\cite
{mole:prb:03,mish:prb:03}.) The possibility to induce spin
accumulation, or a nonequilibrium magnetization, by applying
an electric field only is very interesting from a
basic--science point of view, and would certainly be of great
importance for spintronics applications. In this review, we
show how the interplay between quantum confinement and
spin--orbit coupling in quasi--onedimensional (1D) systems
leads to just such a situation~\cite{uz:prb:02b}.

We are focusing on the electronic structure, transport
properties, and interaction effects in ballistic quantum wires
with Rashba spin splitting present. In the following Section,
we formally introduce the theoretical model under
consideration. The limits of weak and strong spin--orbit
coupling will be distinguished by comparison of the two
fundamental length scales involved, namely the wire width $W$
and the Rashba spin--precession length $L_{\mathrm{so}}$. In
Sec.~\ref{spin_sec}, the spin properties of electronic states
in quasi--1D subbands are revealed. Subband hybridization for
the case of strong spin--orbit coupling turns out to result in
an unusual spin structure where the direction of motion for
electrons at the Fermi wave number basically fixes their spin
state. This is in stark contrast to conventional wires where
states for both spin species exist for either propagation
direction. Quantum wires in the limit of strong spin--orbit
coupling exhibit therefore particularly intriguing transport
properties, discussed in detail in Sec.~\ref{trans_sec} based
on the scattering--theory formalism of mesoscopic electron
transport~\cite{datta}. We find, e.g., that application of an
external voltage can lead to a spin accumulation and
concomitant spin--polarized current flow. Possibilities for
experimental confirmation of our predictions will be
elucidated. The quasi--1D subband structure obtained in
Sec.~\ref{bands_sec} will form the basis for a description of
interacting Rashba--split quantum wires in Sec.~\ref{int_sec},
based on generalizations of the Tomonaga--Luttinger
model~\cite{tom:prog:50,lutt:jmp:63}. Again, the interesting
case is the one with strong spin--orbit coupling where the
unusual spin properties of subband states result in peculiar
properties of spin--sensitive correlation functions. We
formulate our conclusions and give a brief outlook in the
final Sec.~\ref{concl_sec}.

\section{Rashba spin splitting of quasi--1D subbands}
\label{bands_sec}

\begin{figure}
\centerline{\includegraphics[width=12cm]{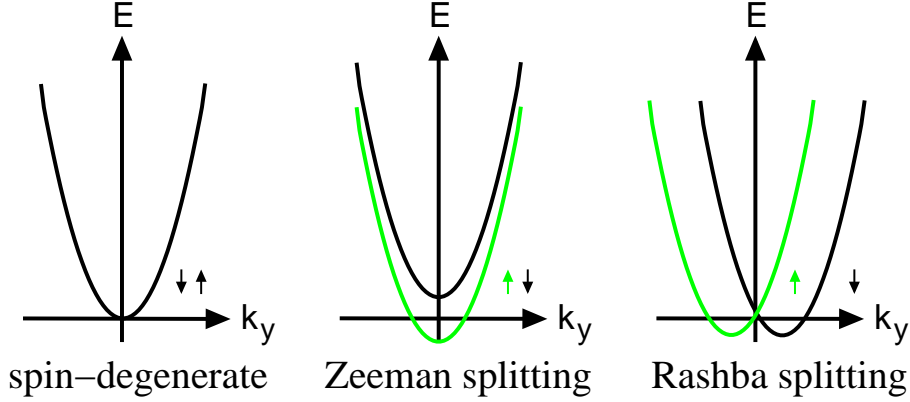}}
\caption{Comparison of Rashba and Zeeman spin splitting. Shown
are energy dispersions for 2D single--electron eigenstates
having a fixed $k_x=0$. Degenerate parabolic curves for both
spin states get to be shifted in {\em energy direction\/} by
an applied magnetic field. Quite differently, finite Rashba
spin splitting results in {\em wave--vector--shifted\/}
dispersion curves.
\label{compsplit}}
\end{figure}

The text--book example of a two--dimensional (2D) quantum well
is now routinely realized by appropriate band--gap engineering
in semiconductor heterostructures~\cite{hetero1}. For low
enough electron densities and temperatures, it is possible to
describe the motion of electrons in the well using the
Hamiltonian
\begin{equation}
H_0=\frac{1}{2 m} \left( p_x^2 + p_y^2 \right)
\end{equation}
of quasi--free particles in the lowest 2D subband. (We take
the growth direction of the heterostructure to be the $z$
axis in our spatial coordinate system.) Corrections to this
Hamiltonian which lead to a coupling of spin state and motion
in real space arise, as in the three--dimensional bulk
material, in the absence of inversion symmetry. In the quantum
well, there exists an additional possibility for breaking
inversion symmetry: creating an asymmetric band bending. For
conduction--band states, the spin--orbit Hamiltonian resulting
from this {\em structural inversion asymmetry\/}~\cite
{rolandbook} has the form~\cite{rashba}
\begin{equation}
H_{\mathrm{so}}=\frac{\hbar k_{\mathrm{so}}}{m} \,\,\left(\vec\sigma
\times \vec p\,\right)\cdot\hat{z}\quad .
\end{equation}
Here $\vec\sigma$ denotes the vector of Pauli matrices, and
the wave number $k_{\mathrm{so}}$ is a direct measure of the
Rashba spin--orbit coupling strength. We use the latter as a
phenomenological input parameter, which has to be determined
experimentally~\cite{nitta:prl:97,schaep:prb-rc:97} or from
spin--dependent electronic--structure calculations~\cite
{andra:prb:94,pfeff:prb:95,silva:165318,rolandbook}. The
single--particle Schr\"odinger equation for the Hamiltonian
$H_{\mathrm{2D}}=H_0+H_{\mathrm{so}}$ describing the 2D
electronic motion can be solved straightforwardly~\cite
{byra:jetplett:84}. Electronic eigenstates are labeled by a
2D wave vector $\vec k$ and the quantum number $\sigma=\pm 1$
of spin projection in the direction perpendicular to both
$\vec k$ and the growth direction. The energy eigenvalues for
states having the same $k=|\vec k|$ but opposite $\sigma$ turn
out to differ by a zero--field spin splitting $\Delta E_k=
\hbar^2 k_{\mathrm{so}} k / m$. Unlike Zeeman splitting, to
which it is often compared, Rashba spin splitting does not
result in a finite global magnetization of the 2D electron
system, as time--reversal symmetry is preserved by the
Hamiltonian $H_{\mathrm{2D}}$. No common spin quantization
axis can be found for its eigenstates. Furthermore, a cut
through the energy dispersions reveals that spin bands are
shifted not by a fixed energy, as is the case for Zeeman
splitting, but rather in wave--vector direction. This is
illustrated in Fig.~\ref{compsplit}.

\begin{figure}
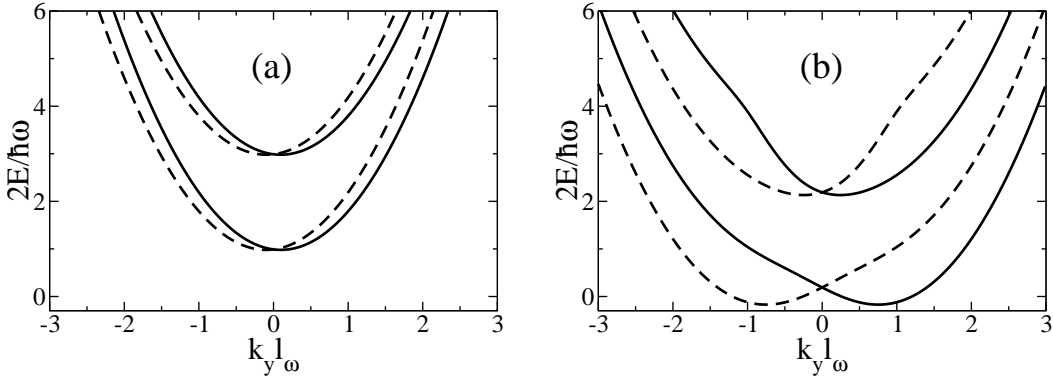

\includegraphics[height=5cm]{fig2a.eps}
\hfill\includegraphics[height=5cm]{fig2b.eps}
\caption{Quantum--wire subband dispersions in the limit of
weak and strong Rashba spin splitting. (a)~Parabolic
dispersions shifted in wave--vector direction arise for weak
spin splitting when the wire width is much smaller than the
spin precession length. Each subband can be associated with a
good quantum number of spin projection in the direction
perpendicular to the wire. Data shown are for $\ell_\omega
k_{\mathrm{so}}=0.1$. (b)~In the opposite case of strong spin
splitting, hybridization between adjacent subbands with
opposite spin results in a nonparabolic dispersion. We show
data for the case $\ell_\omega k_{\mathrm{so}}=0.9$.
\label{displimit}}
\end{figure}

In the following, we are focusing exclusively on the situation
where the motion of electrons is further confined to one
spatial dimension by an external potential $V(x)$. To be
specific, we assume a parabolic confinement
\begin{equation}
V(x)=\frac{m}{2} \,\,\omega^2 \, x^2 \quad ,
\end{equation}
as we can expect the qualitative features of spin splitting in
quantum wires to be independent of the actual shape of the
confining potential. In the absence of spin--orbit coupling,
the single--electron spectrum is split into quasi--1D subbands
having quadratic dispersion in the 1D wave vector $k_y$ that
labels its eigenstates. The characteristic energy scale for
subband bottoms is related to the width $W$ of the quantum
wire. For finite $k_{\mathrm{so}}$, energy eigenstates are
still plane waves in wire direction, but the linear dependence
of $H_{\mathrm{so}}$ on the momentum $p_x$ in confinement
direction introduces a coupling between the quasi--1D
subbands. Using the 1D plane--wave representation, we can
write the quantum--wire Hamiltonian in the form
$H_{\mathrm{qw}}=H_{\mathrm{sb}}+H_{\mathrm{mix}}+
H_{\mathrm{1D}}$, where
\begin{eqnarray}
H_{\mathrm{sb}} &=& \frac{p_x^2}{2 m} + V(x) \quad ,\\
H_{\mathrm{mix}} &=& - \frac{\hbar k_{\mathrm{so}}}{m} \,\,
\sigma_y \, p_x \quad ,\\
H_{\mathrm{1D}} &=& \frac{\hbar^2}{2 m} \left( k_y +
k_{\mathrm{so}} \sigma_x \right)^2 - \frac{\hbar^2
k_{\mathrm{so}}^2}{2 m} \quad .
\end{eqnarray}
The importance of quantum--wire subband coupling can be
quantified by the ratio $s$ of the matrix elements of
$H_{\mathrm{mix}}$ between eigenstates of $H_{\mathrm{sb}}+
H_{\mathrm{1D}}$ (which describes a hypothetical quantum wire
having only $p_y$--dependent spin splitting) and the
difference of the corresponding eigen energies~\cite
{mire:prb:01}. Estimating the parameter $s$ in terms of the
wire width $W$, we find
\begin{equation}\label{weakstrong}
s\approx \frac{\hbar k_{\mathrm{so}}}{m}\,\frac{\pi\hbar}{W}
\left(\frac{\hbar^2 \pi^2}{m W^2}\right)^{-1} = \frac{W
k_{\mathrm{so}}}{\pi}\equiv \frac{W}{L_{\mathrm{so}}}
\quad ,
\end{equation}
where $L_{\mathrm{so}}$ is the spin precession length familiar
from the proposed spinFET~\cite{spinfet}. Hence two limits can
be distinguished. (i)~Quantum wire with weak Rashba spin
splitting, realized for $W\ll L_{\mathrm{so}}$: Eigenstates
are plane--wave spinors that are eigenstates of $\sigma_x$,
i.e., have their spin polarized in the direction perpendicular
to (and in the plane of) the wire. The energy dispersion is
given by two parabolas, distinguished by the spin quantum
number, that are shifted in wave--vector direction by an
amount $2 k_{\mathrm{so}}$. This is illustrated in
Fig.~\ref{displimit}a, which is essentially identical to the
cut through a 2D Rashba--split dispersion shown in Fig.~\ref
{compsplit}. (ii)~Quantum wire with strong Rashba spin
splitting, realized when $W\gtrsim L_{\mathrm{so}}$:
Hybridization of quasi--1D subbands results in a nonparabolic
energy dispersion~\cite{moroz:prb:99,mire:prb:01}. See
Fig.~\ref{displimit}b. In this case, no common spin quantum
number can be assigned to states within a spin--split
quantum--wire subband~\cite{uz:prb:02b}. Their intriguing spin
properties will be discussed in the following Section.

A useful analytical result can be obtained for energy
eigenvalues of states in a parabolic quantum wire having
$k_y=0$. As $H_{\mathrm{1D}}$ vanishes in this case, these are
eigenspinors of $\sigma_y$ whose components are eigenfunctions
of harmonic oscillators with a spin--dependent boost. The
energy eigenvalue is degenerate in the good quantum number of
spin projection in the wire direction. With $l_\omega=\sqrt
{\hbar/(m\omega)}$ denoting the oscillator length scale
introduced by the parabolic wire confinement, it is explicitly
given by
\begin{equation}\label{specstate}
E_n(k_y=0) = \frac{\hbar\omega}{2}\left[2 n + 1 - (\ell_\omega
k_{\mathrm{so}})^2\right] \quad . 
\end{equation}
The spin--orbit correction to the energy of these states
arises due to their finite quantized motion in the direction
perpendicular to the wire. This exact result is an important
benchmark to judge the accuracy of numerical methods that are
necessary to obtain the full dispersion curves for all wave
vectors~\cite{moroznote}.

\section{Spin properties of quantum--wire eigenstates}
\label{spin_sec}

While it is not possible to find a common spin quantization
axis for eigenstates in a 2D system with Rashba spin
splitting, restriction to one spatial propagation direction
restores a global spin--projection axis {\em for the case of
weak spin--orbit coupling}. In this limit, it is possible to
neglect $H_{\mathrm{mix}}$, and the single--electron states in
the wire are given, to a good approximation, by eigenstates of
spin projection parallel to the wire confinement, i.e., the
$x$ direction. At any given energy, there exist two such
eigenstates with opposite spin, which have wave vectors
differing by $2 k_{\mathrm{so}}$. The effect of finite
$H_{\mathrm{mix}}$ is to modify the local spin density of
eigenstates across the wire, introducing a small zero--average
tilt out of the wire plane~\cite{haeus:prb-rc:01}. This
inhomogeneity of spin density as a function of the coordinate
perpendicular to the wire becomes more important for strong
spin--orbit coupling~\cite{uz:prb:02b}. Still, direct
experimental observation of this texture--like spin structure
would be quite challenging.

The mixing term $H_{\mathrm{mix}}$ couples eigenstates
obtained from diagonalizing $H_{\mathrm{sb}}+H_{\mathrm{1D}}$
which have opposite spin. In the special case of a parabolic
confinement, its matrix elements turn out to be finite only
between those opposite--spin states whose oscillator--band
indices differ by one~\cite{genconf}. Hence, wherever the
dispersion relations for such energetically adjacent subbands
with opposite spin cross, $H_{\mathrm{mix}}$ cannot be
neglected, as it will induce an anticrossing. Another way to
define the limit of weak spin--orbit coupling is then to say
that such anticrossings occur only at energies that are much
higher than the typical quasi--1D subband splitting and,
hence, can be neglected in the case when only a few wire
subbands are occupied. As the parameter $s$ defined in
Eq.~(\ref{weakstrong}) gets close to unity, anticrossings
occur at energies comparable to those of low--lying wire
subbands and affect its properties in an important way.
Following the picture of anticrossings, it becomes immediately
apparent that a peculiar spin structure emerges for states in
hybridized subbands. In particular, their eigen--spin
direction becomes a function of wave vector and is not uniform
anymore within each band.
\begin{figure}
\centerline{\includegraphics[height=5cm]{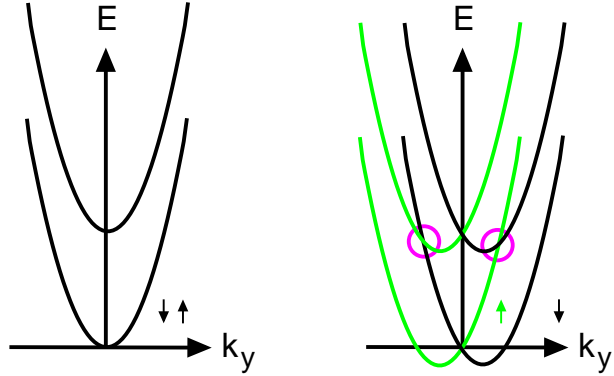}}
\caption{Effect of subband mixing in the limit of strong
spin--orbit coupling. Shown in the left panel are two
spin--degenerate subbands in the absence of Rashba spin
splitting. Dispersions obtained from diagonalizing $H_{\mathrm
{sb}}+H_{\mathrm{1D}}$ are parabolas shifted, due to
Rashba spin splitting, in wave--vector direction, as shown
in the right panel. Corresponding eigenstates have spin
polarized in the direction perpendicular to the wire. The
mixing Hamiltonian $H_{\mathrm{mix}}$ induces subband
hybridization near crossing points, indicated on the right
panel by circles. As a result, states in the lowest two
spin--split subbands have approximately parallel spin at large
wave vectors.
\label{anticross}}
\end{figure}

\begin{figure}
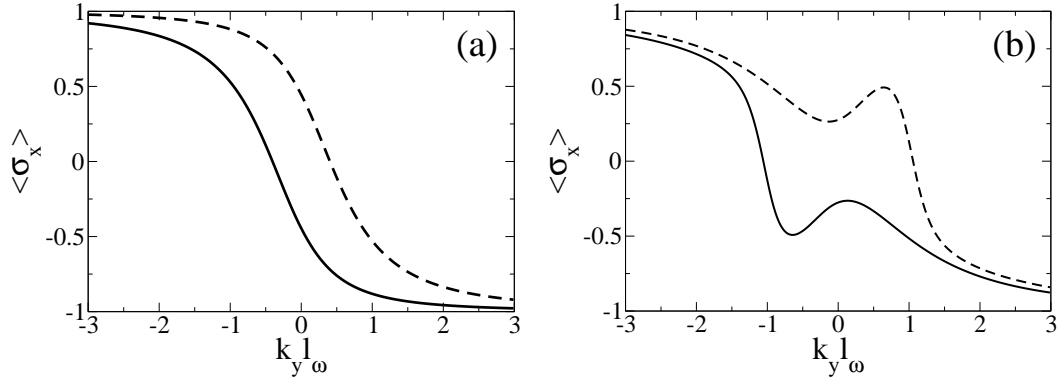

\includegraphics[height=5cm]{fig4a.eps}\hfill
\includegraphics[height=5cm]{fig4b.eps}
\caption{Spin projection of eigenstates in the lowest (a) and
first excited (b) spin--split subbands for the situation when
$\ell_\omega k_{\mathrm{so}}=0.9$. (See the corresponding
dispersions in Fig.~\ref{displimit}.) In both cases, states at
large wave vector have approximately parallel spin. 
\label{spinpic}}
\end{figure}

Consider the hybridization of the lowest two spin--split
subbands, as illustrated in Fig.~\ref{anticross}. The
resulting nonparabolic dispersion for a particular parameter
$l_\omega k_{\mathrm{so}}$ is given in Fig.~\ref{displimit}b.
Note that states in the lowest pair of subbands exist with
wave vector larger than that for the crossing point but
energies smaller than the bottoms of the next subband pair.
These states originated from the first--excited spin--down
subband but have become part of the lowest subband after
hybridization. Their wave--vector distance to the crossing
point is large enough such that the spin--up admixture in the
eigenspinors is quite small. Hence we find a situation where
right--moving states at still comfortably low energies in the
lowest spin--split subbands have basically parallel spin! For
the case depicted, there are basically only spin--down
right--movers at large enough energies, and similarly only
spin--up left--movers. This direct association of spin state
with the direction of motion is peculiar to the
strong--Rashba--split situation. In an ordinary quantum wire,
even with Zeeman splitting or weak Rashba splitting present,
there exist right--moving states for both spin--up and
spin--down electrons {\em at any energy\/}. It can be
envisioned that, for suitable parameter ranges, there exists a
finite energy window for which even more than two sets of
low--lying subbands have right--moving (and left--moving)
states with almost parallel spin. This turns out to indeed be
the case for $\ell_\omega k_{\mathrm{so}}=0.9$. In
Fig.~\ref{spinpic}, we show the expectation value of the spin
projection perpendicular to the wire for eigenstates in the
lowest two pairs of spin--split quantum--wire subbands.
In both sets of subbands, right--moving states at large wave
vector exist only for spin--down electrons, and left--moving
ones only for spin--up~\cite{correctnote}.

Exactly at the anticrossing, eigenstates of $H_{\mathrm{qw}}$
are the symmetric and antisymmetric superpositions of spin--up
and spin--down states from the two intersecting quasi--1D
subbands for $H_{\mathrm{sb}}+H_{\mathrm{1D}}$. Due to this
peculiar mixing of spin and subband wave functions,
expectation values of spin projected in any direction yields
zero for quantum--wire eigenstates at the anticrossing. Their
special spin properties turn out to give rise to an additional
spin rotation of incoming electrons with energies within the
anticrossing gap~\cite{eguesprl}, which would enable enhanced
performance of a suitably designed spinFET~\cite{eguesapl}.

\section{Spin--dependent transport from electric fields only}
\label{trans_sec}

The key theoretical tool for studying linear transport in
mesoscopic systems is the Landauer--B\"uttiker formalism~\cite
{datta}, which is valid when electron--electron interaction is
negligible. The central result of this approach is that the
conductance of a mesoscopic sample can be related to the
scattering matrix, $\mathcal{S}$, of the sample.  In the
spin--independent case, the scattering matrix is diagonal in
spin space, and the problem can be mapped to scattering of
spinless particles. Spin is then taken into account only as a
factor $2$, and the linear conductance at zero temperature
simply reads $G=\frac{2e^2}{h}\sum_{j\in 1,i\in 2} |\mathcal
{S}_{i, j}|^2$, where $\mathcal{S}_{i, j}$ is the scattering
amplitude (transmission coefficient) between the mode $j$ in 
terminal $1$ and the mode $i$ in terminal $2$. In this
situation (spin--independent case),  time--reversal symmetry
implies $\mathcal{S}=\mathcal{S}^{\mathrm{T}}$, i.e.,
$\mathcal{S}_{i,j}=\mathcal{S}_{j,i}$, where, in the generic
multi-terminal case,  $i$ and $j$ are collective indices
labeling both terminals and transverse modes. In the presence
of spin--orbit coupling, there is no common spin quantization
axis, and the full spin structure of $\mathcal{S}$ should be
retained~\cite{feve}. The total two-terminal conductance reads
then $G=\frac{e^2}{h}\sum_{j\in 1,i\in 2,\sigma,\sigma^\prime}
|\mathcal{S}_{i\sigma, j\sigma^\prime}|^2$, where $\sigma$ and
$\sigma^\prime$ denote the spin projection along a common
quantization axis. The spin--$\sigma$ current in the output
contact $2$ is given by $I_\sigma=G_\sigma V$, where
\begin{equation}
\label{gsigma}
G_\sigma=  \frac{e^2}{h}\sum_{j\in 1,i\in 2,\sigma^\prime} 
|\mathcal{S}_{i\sigma, j\sigma^\prime}|^2 \quad . 
\end{equation}
In Eq.~(\ref{gsigma}) the sum over the transmission
probabilities for different incoming spin projections
$\sigma^\prime$ describes the fact that the injecting
reservoir $1$ is not spin--polarized. In this case of a
nondiagonal scattering matrix in spin space, the condition
satisfied by the scattering matrix due to time--reversal
symmetry reads
\begin{equation}\label{smat_cond}
\mathcal{S}_{i\sigma,j\sigma^\prime}= (\sigma\sigma^\prime)
\mathcal{S}_{j-\sigma^\prime, i-\sigma},
\end{equation}
where $\sigma$ and $\sigma^\prime$ denote the spin projection
along a common quantization axis and can take the values $\pm
1$. The condition Eq.~(\ref{smat_cond}) does not forbid the
creation of spin-polarized currents in linear transport. 

\begin{figure}
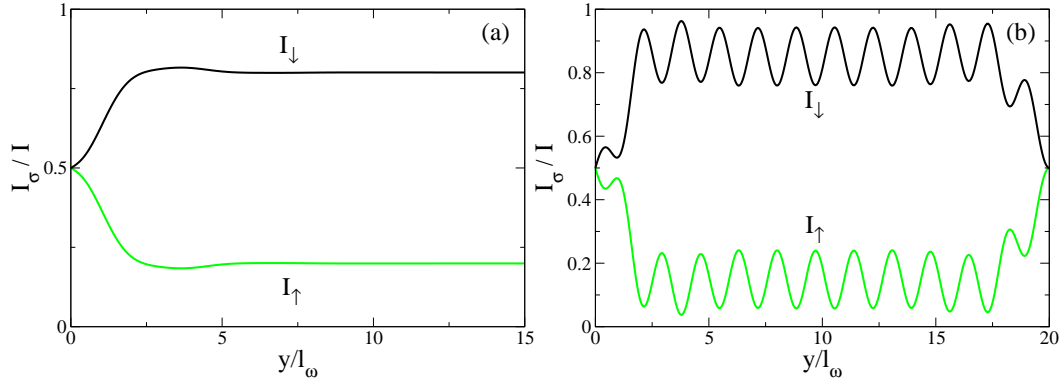

\includegraphics[height=5cm]{fig5a.eps}\hfill
\includegraphics[height=5cm]{fig5b.eps}
\caption{Transport in the hybrid wire/leads system computed by
means of Landauer--B\"uttiker formalism. The total
spin-$\sigma$ current $I\sigma$ (where the quantization axis
is along $x$) is plotted as a function of the coordinate along
the wire. In panel (a) there is only one interface between a
semi--infinite lead ($y<0$) and a semi--infinite wire with
spin-orbit coupling. In panel (b) the wire is contacted by two
semi--infinite leads. The two interfaces with the leads are
located at $y=0$ and $y=20\ell_\omega $, respectively. In both
cases, reservoirs inject spin-unpolarized electrons. The Fermi
energy is chosen such that only one subband is occupied. The
oscillatory behavior seen in panel (b) is due to quantum
interference caused by multiple scattering between the
interfaces. The parameters used in the simulation are
$E_{\mathrm{F}}=\hbar \omega$ and $\ell_\omega k_{\mathrm{so}}
=0.9$. The calculation has been performed within the two--band
model described in Ref.~\cite{uz:prb:02b}.
\label{curr}}
\end{figure}
We now consider a wire with Rashba spin-orbit coupling with
two nonmagnetic contacts. These are described by semi-infinite
leads with no spin--orbit coupling. In the Landauer-B\"uttiker
approach, right--movers are populated by the left reservoirs
(with chemical potential $\mu_{\mathrm{L}}$) and left--movers
by the right reservoirs (with chemical potential $\mu_{\mathrm
{R}}$). This consideration, together with the observation that
we can realize situations in which the chirality (direction of
motion) fixes the spin state (as discussed in the previous
section), leads to the expectation of spin accumulation in the
wire when a transport voltage is applied.

\begin{figure}
\includegraphics[height=5cm]{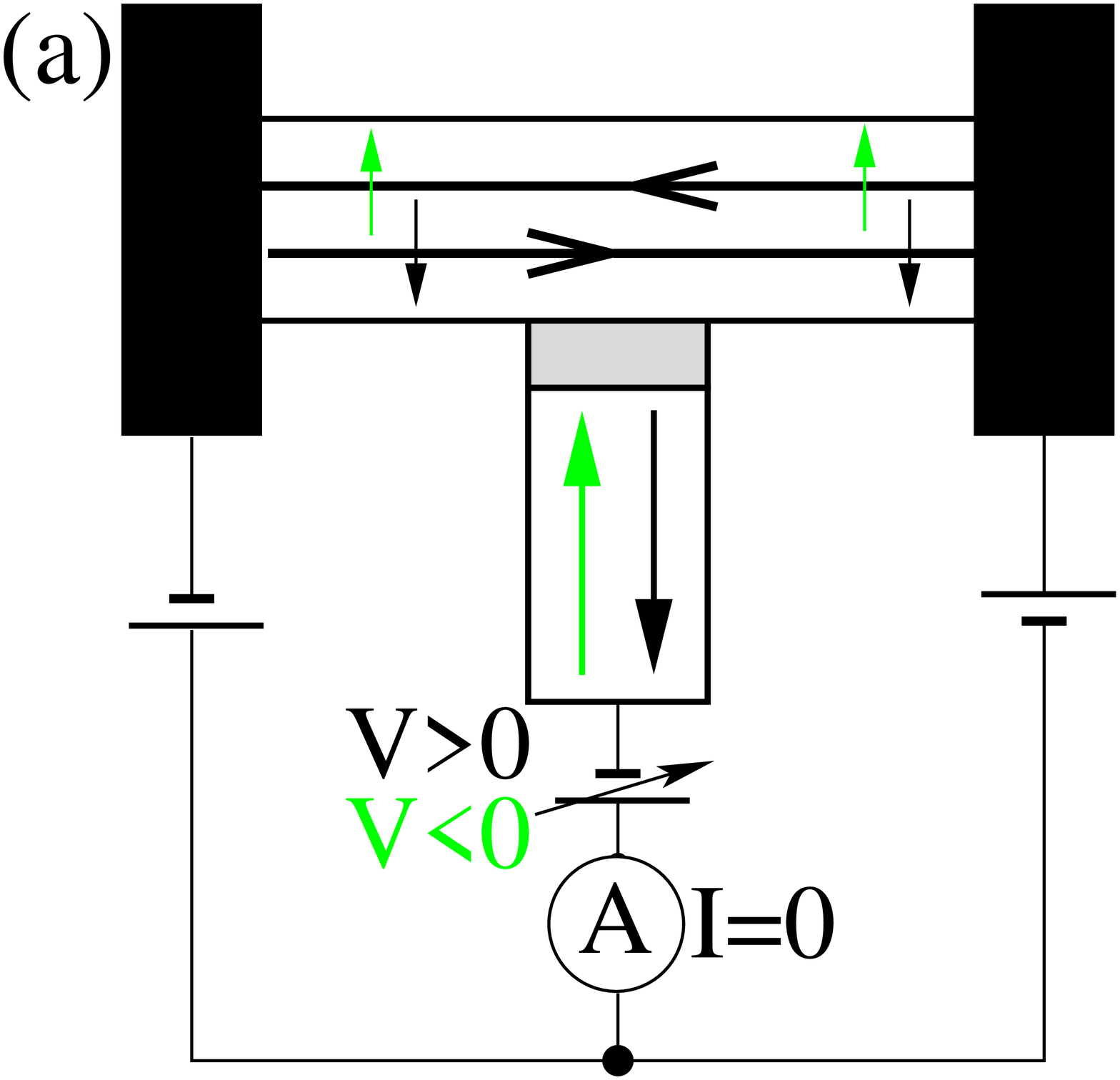}\hfill
\includegraphics[height=5cm]{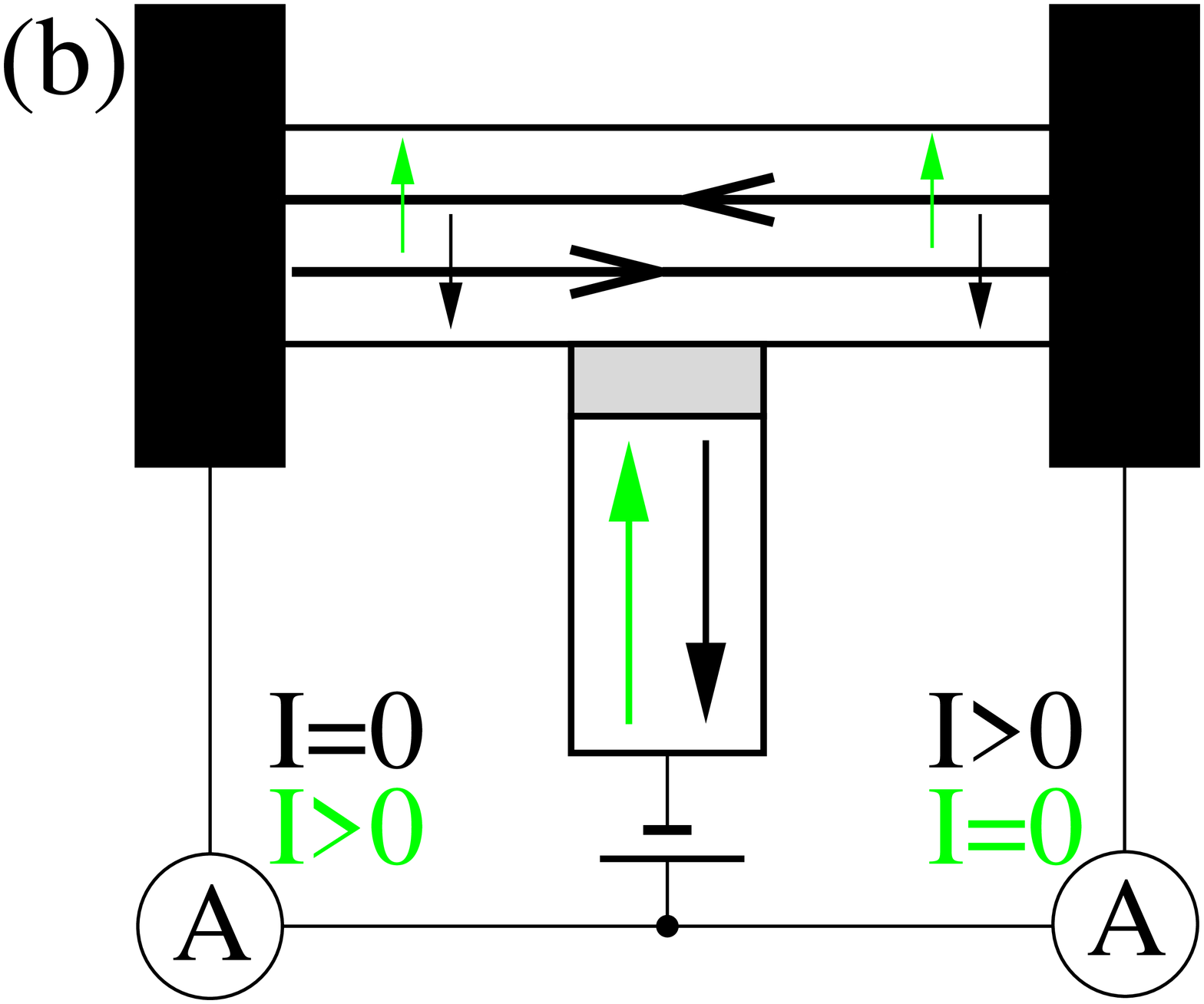}
\caption{Schematic description of suggested experimental
setups for verification of spin accumulation. Panel~(a):
Measurement of spin--dependent electrochemical potential with
ferromagnetic voltage probe. Panel (b): Injection of
spin--polarized current via tunnel barrier from ferromagnetic
contact, and subsequent measurement of the {\em
directionality\/} of current flow probes the chirality of the
allowed spin states. Depending on the direction of the
magnetization, only one of the two Amperemeters detects a
finite current~\cite{meijer}. 
\label{experiment}}
\end{figure}
To obtain linear--transport currents, the scattering problem
is solved by standard mode--matching techniques. The usual
condition for the probability--current conservation needs to
be modified because the velocity operator in the presence of
Rashba spin-orbit coupling
reads~\cite{mol:prb-rc:01,uz:prl:02}
\begin{equation}\label{velop}
v_y =-\frac{i}{\hbar}[y,H_{\mathrm{qw}}]= \hbar (k_y +
k_{\mathrm{so}}\sigma_x)/m \quad .
\end{equation}
To have a clear understanding of what happens at the interface
between a lead without spin--orbit coupling and the wire with
spin--orbit coupling, we study first the simple situation of a
single interface between a semi--infinite lead and a
semi--infinite wire with spin--orbit coupling. The results for
the  total spin-up(-down) current as a function of the
distance from the interface are shown in Fig.~\ref{curr}a. The
current flowing in the lead is unpolarized, while in the wire,
distant enough from the interface, the current becomes
spin--polarized in agreement with the spin properties of the
eigenstates shown in Fig.~\ref{spinpic}a. Hence, the Rashba
spin splitting leads to \textit{spin accumulation} in the wire
without ferromagnetic contacts. The other important phenomenon
which is apparent from Fig.~\ref{curr} is the process of
\textit{current conversion} occurring at the interface. This
process is mediated by scattering into evanescent modes of the
wire, and is possible thanks to the anomalous form of the
velocity operator (\ref{velop}). Fig.~\ref{spinpic}b shows
results for the experimentally relevant situation of a
finite--length wire with spin--orbit coupling  between two
leads. Also in this case, we have both spin accumulation and  
current conversion, plus some finite-size oscillation due 
to quantum interference~\cite{subbands}. 

Finally we suggest two possible schemes to enable experimental
verification of spin accumulation (see Fig.~\ref{experiment}).
One possibility would be to weakly couple a ferromagnetic
voltage probe to the wire, measuring the chemical potential of
the majority spins. Changing the direction of the
magnetization of the ferromagnetic contact enables the
detection of the difference in chemical potentials of the two 
spin species. Another possibility could be to use injection of
a spin--polarized current from a ferromagnet to probe the
chirality of propagating spins  in the wire. The direction of 
current flow should then depend on the direction of the 
magnetization of the ferromagnet~\cite{meijer}.

\section{Electron--electron interactions and Rashba spin
splitting}
\label{int_sec}

Interactions between electrons turn out to have more dramatic
consequences in 1D as compared to higher--dimensional systems.
Instead of following the familiar~\cite{pines} Fermi--liquid
paradigm, 1D conductors form their own new class of
interacting metals dubbed {\em Luttinger liquids\/}~\cite
{fdmh:jpc:81,voit:reprog:94}. Power--law behavior of
electronic correlation functions, such as the tunneling
density of states or the momentum--space occupation number,
near the Fermi points are a signature of non--Fermi--liquid
properties exhibited by interacting 1D systems. Previous
work~\cite{haeus:prb-rc:01} has shown that weak Rashba spin
splitting leaves such power laws essentially unaffected.
However, the interplay between Luttinger--liquid behavior and
spin--orbit coupling turns out to be nontrivial in quantum
wires with {\em strong\/} Rashba spin splitting and deserves a
more detailed discussion.

\begin{figure}
\includegraphics[height=4.3cm]{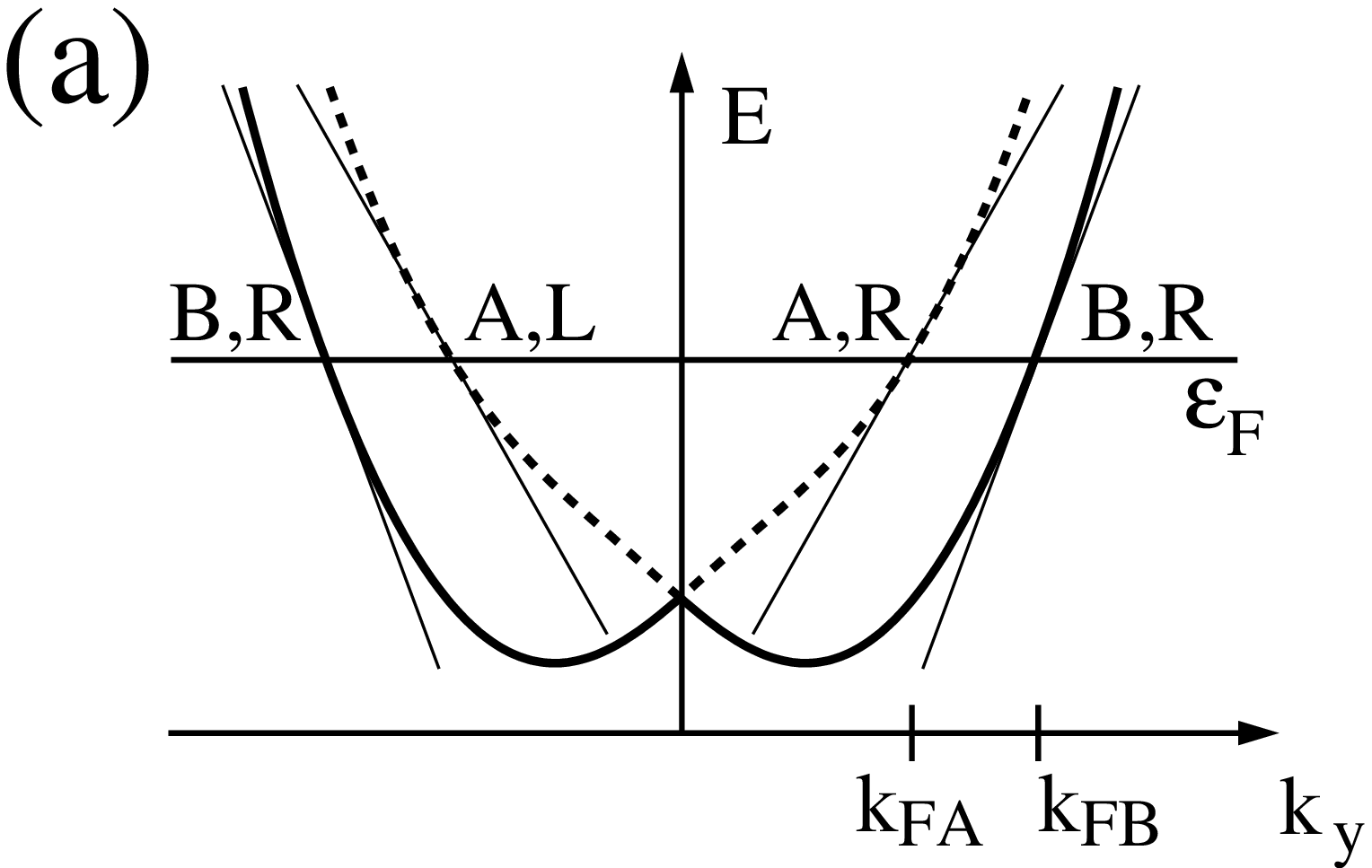}\hfill
\includegraphics[height=4.3cm]{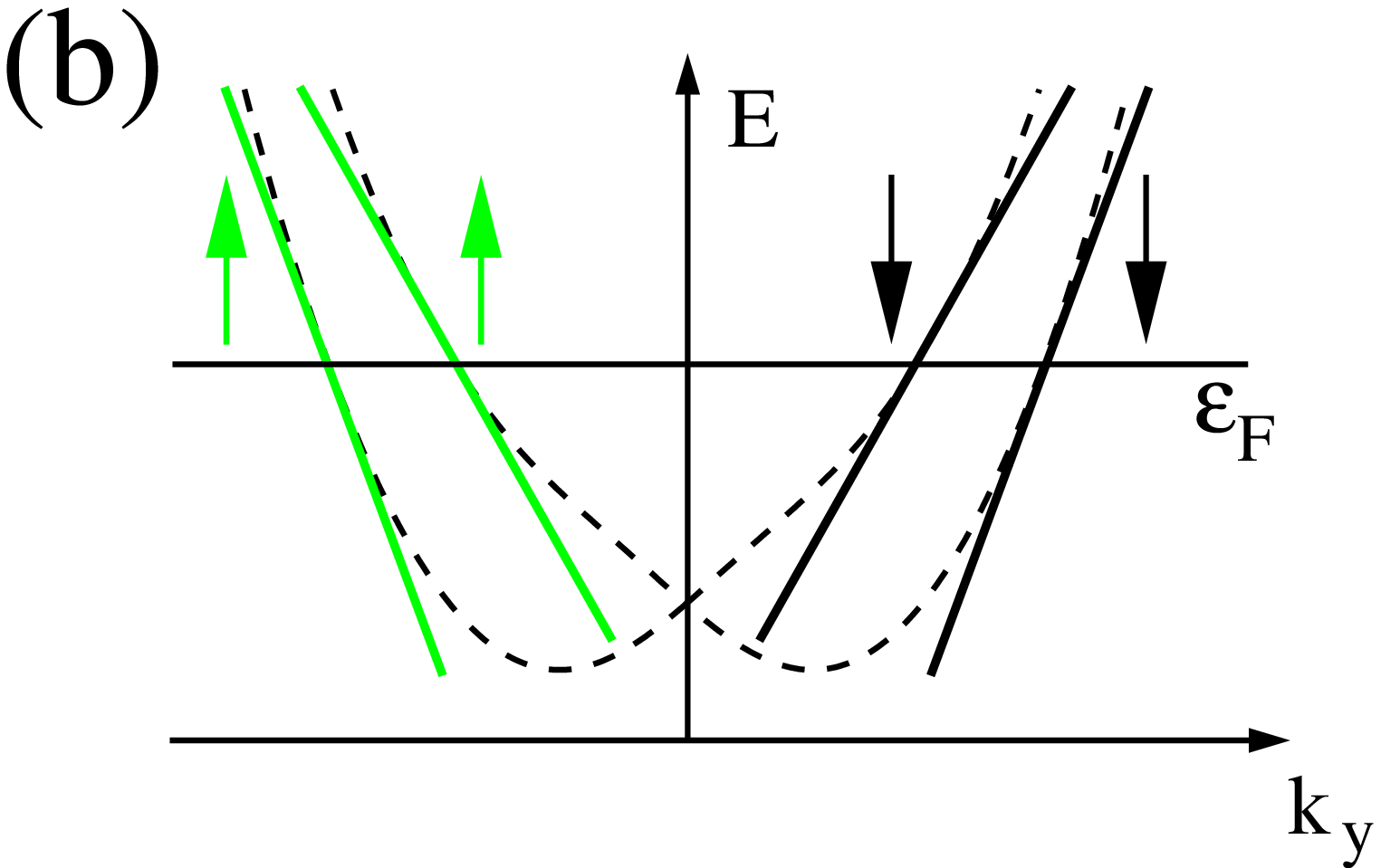}
\caption{Linearization of the quasi--1D spin--split subband
energy dispersion is the starting point for application of the
Tomonaga--Luttinger model. (a)~In general, we can distinguish
two branches of right--movers (R) and left--movers (L) each,
labeled A and B here. Their most conspicuous differences are
in the magnitude of their Fermi wave numbers and velocities.
When Fermi points are located far enough from anticrossings,
the situation depicted in (b) arises where electrons from both
right--moving branches are in the same spin state. Similarly,
left--moving states then turn out to have parallel spin, which
is opposite to that of the right--movers
\label{lineardisp}}
\end{figure}

The current understanding of interaction effects in 1D systems
is based on the possibility to map real systems onto the
exactly soluble Tomonaga--Luttinger model~\cite
{tom:prog:50,lutt:jmp:63}. The starting point of such a
mapping is linearization of the single--particle dispersion
relation near the Fermi points. Furthermore, two chiral
electron flavors (right--movers and left--movers) are
distinguished according to their velocity direction. In the
present case of a strong Rashba--split quantum wire, the spin
degeneracy of these chiral electron branches is broken. In the
most general case, we can distinguish two right--moving
branches having different velocities. As discussed in
Sec.~\ref{spin_sec} above, their spin properties are peculiar
because they result from hybridization of quasi--1D subbands
with opposite spin. In particular, when Fermi points are
sufficiently far away from anticrossings, both right--moving 
branches can be modeled as having approximately parallel spin.
The same is then true for left--movers which, however, have
their spin opposite to that of right--movers. To be most
general, we do not {\it a priori\/} assign spin labels to the
chiral electron branches. Instead, we distinguish them by
their different velocities. See Fig.~\ref{lineardisp} for an
illustration.
The single--electron part of the Tomonaga--Luttinger
Hamiltonian of a quantum wire with strong Rashba spin
splitting can then be written as
\begin{equation}
H_{\mathrm{TL}}^{\mathrm{(0)}}= \sum_{\alpha={\mathrm{A,B}}}
\sum_k \hbar v_\alpha k \left(c_{\mathrm{R}\alpha}^\dagger
c_{\mathrm{R}\alpha} - c_{\mathrm{L}\alpha}^\dagger
c_{\mathrm{L}\alpha}\right) \quad .
\end{equation}
We have omitted uniform shifts of the chemical potential and
total energy of the system. Introducing the familiar chiral
bosonic phase fields~\cite{fdmh:jpc:81,voit:reprog:94}
$\phi_{\nu\alpha}$ that are related to the normal--ordered
electron density for each branch via $\varrho_{\nu\alpha}=
\partial_x\phi_{\nu\alpha}/(2\pi)$, with $\alpha=\mathrm{A,B}$
and $\nu=\mathrm{R,L}$, we can write the single--particle and
interaction parts of the Tomonaga--Luttinger Hamiltonian
$H_{\mathrm{TL}}=H_{\mathrm{TL}}^{\mathrm{(0)}}+
H_{\mathrm{TL}}^{\mathrm{(int)}}$ as
\begin{eqnarray}
H_{\mathrm{TL}}^{\mathrm{(0)}}&=&\frac{\hbar}{4\pi}\int d x
\sum_{\nu\alpha} v_\alpha \left(\partial_x \phi_{\nu\alpha}
\right)^2 \quad , \\
H_{\mathrm{TL}}^{\mathrm{(int)}}&=&\frac{V_0}{8\pi^2}\int d x
\left(\sum_{\nu\alpha}\partial_x\phi_{\nu\alpha}\right)^2
\quad .
\end{eqnarray}
Here we have only included the long--wave--length
forward--scattering component $V_0$ of the screened Coulomb
interaction in the wire. Changing representation to the
nonchiral conjugate phase fields $\theta_\pm=(\sum_\nu
\phi_{\nu\mathrm{B}}\pm\sum_\nu\phi_{\nu\mathrm{A}})/\sqrt{8
\pi}$ and $\Pi_\pm=(\partial_x\phi_{\mathrm{R}\mathrm{B}}-
\partial_x\phi_{\mathrm{L}\mathrm{B}}\pm[\partial_x
\phi_{\mathrm{R}\mathrm{A}}-\partial_x\phi_{\mathrm{L}\mathrm
{A}}])/\sqrt{8\pi}$, and introducing the abbreviations $v_+=(
v_{\mathrm A}+v_{\mathrm B})/2$, $K_\rho=\sqrt{1+\frac{2 V_0}
{\pi v_+}}$, $v_{\mathrm{pl}}=v_+/K_\rho$, and $\delta=(
v_{\mathrm B}-v_{\mathrm A})/v_+$, we find after some algebra
$H_{\mathrm{TL}}=H_+ + H_- + H_\delta$, where
\begin{eqnarray}
H_+&=&\frac{\hbar v_{\mathrm{pl}}}{2}\int d x\left\{ \frac{1}
{K_\rho} (\partial_x\theta_+)^2 + K_\rho \, \Pi_+^2\right\}
\quad , \\
H_-&=&\frac{\hbar v_+}{2}\int d x\left\{ (\partial_x\theta_-)^2 +
\Pi_-^2\right\} \quad , \\
H_\delta&=&\delta\, \frac{\hbar v_+}{2}\int d x\left\{(\partial_x
\theta_+)(\partial_x\theta_-) + \Pi_+ \, \Pi_- \right\}\quad .
\end{eqnarray}
Comparison with the Tomonaga--Luttinger model Hamiltonian of
an ordinary quantum wire~\cite{voit:reprog:94} shows that
$H_+$ describes plasmon excitations, i.e., fluctuations in the
total electron density of the wire, where $K_\rho$ is the
usual Luttinger parameter. The term $H_-$ represents
long--wave--length excitations in the density difference
between electrons (both right--moving and left--moving) in the
B and A branches, which is unaffected by Coulomb interactions.
Note that $\theta_-$ and its conjugate  field $\Pi_-$ {\em do
not\/} represent the spin density in the strongly
Rashba--split quantum wire --- not even in the limit where
electrons close to the Fermi points have approximately
parallel spin, as depicted in Fig.~\ref{lineardisp}b. Finally,
$H_\delta$ arises because of the velocity difference for
electrons from the A and B branches. Its existence leads to
the emergence of four different chiral normal modes that
diagonalize $H_{\mathrm{TL}}$, in contrast to the case of an
ordinary quantum wire where pairs of left--moving and
right--moving normal modes with equal velocities exist that
can be combined to two nonchiral normal modes. (The latter
represent total--charge and total--spin--density waves,
respectively.)

The peculiar spin structure of single--electron states in
strongly Rashba--split subbands gives rise to an unusual
expression for the Zeeman energy in an external magnetic field
$B$. Focusing on the case depicted in Fig.~\ref{lineardisp}b,
a straightforward calculation yields
\begin{equation}
H_{\mathrm Z} = \frac{g_{\mathrm e}\mu_{\mathrm B}B}{\sqrt{2
\pi}}\int d x \,\, \Pi_+ \quad ,
\end{equation}
with electron g--factor $g_{\mathrm e}$ and Bohr magneton
$\mu_{\mathrm B}$.

The above expressions for $H_{\mathrm{TL}}$ and $H_{\mathrm Z}
$ are quadratic in the bosonic phase fields $\theta_\pm$ and
their conjugates $\Pi_\pm$. Together with the bosonization
identities~\cite{vondelft} for electrons from each of the
chiral branches, it is possible to obtain exact results for
electronic correlation functions, which are omitted here
because of limited space. In the absence of an external
magnetic field, spin--independent correlation functions are
identical to those of a two--component Luttinger liquid
discussed, e.g., in Refs.~\cite{penc:prb:93,aoki:prb:96}.

\section{Conclusions and outlook}
\label{concl_sec}

Our study of Rashba spin splitting in low dimensions has
given us new insight into the interplay between quantum
confinement and spin--orbit coupling. Intriguing
spin--dependent transport effects arise in quantum wires
having a width that is comparable to the spin--precession
length. These could form the basis for realizing spintronics
devices without involving any magnetic parts. Possibilities
for experimental verification of our predictions have been
suggested, which could be realized using present
nanofabrication techniques. It will be interesting to
investigate, both experimentally and theoretically,
spin--orbit effects in related quantum--confined structures
such as quantum point contacts and non--semiconductor--based
systems.

\section*{Acknowledgment}

This work was supported in part by the Center for Functional
Nanostructures at the University of Karlsruhe, Germany. We
benefitted from several useful discussions with F.~E.~Meijer
and A.~F.~Morpurgo.





\begin{thebibliography}{10}
\bibitem{moroznote}
We suspect that the shift of quasi--1D Rashba--split subbands
to lower energies, as apparent from Eq.~(\ref{specstate}),
actually prevents the appearance of `bumps' in the dispersion
relation that were claimed to exist in Ref.~\cite
{moroz:prb:99}. See Fig.~5c in that paper which shows their
numerical results for $l_\omega k_{\mathrm{so}}=0.9$. For the
purpose of direct comparison, we give our numerical results
for the same parameter in our Fig.~\ref{displimit}b. We find
no bumps. Surprisingly, the energies for states at $k_y=0$
obtained numerically in Ref.~\cite {moroz:prb:99} are
unshifted from the subband--bottom energy at zero $k_{\mathrm
{so}}$, in disagreement with the exact analytical result of
our Eq.~(\ref {specstate}).
\bibitem{genconf}
For a general shape of the confinement potential, mixing is
finite between any two subbands with opposite spin, but it is
always largest for those adjacent in energy.
\bibitem{correctnote}
The spin properties of right--movers and left--movers in
strong--Rashba--split quantum wires were presented incorrectly
in previous works by Moroz and Barnes~\cite
{moroz:prl:00,moroz:prb:00b}, which has mislead subsequent
related studies~\cite {egg:epl:01,iucci,yuetal}.
\bibitem{subbands}
The results shown in Fig.~\ref{curr} differ from those
presented in Fig.~4 of  Ref.~\cite{uz:prb:02b} because, when
calculating the latter, the confinement in the leads was
taken to be different from that in the wire. (For
data shown  in Fig.~4 of Ref.~\cite{uz:prb:02b}, the first two
subband energies  in the leads were $(3/4) \hbar\omega$ and
$(5/4) \hbar\omega$, respectively.) Data shown here are for a
uniform confinement potential in the entire structure
consisting of the leads and the Rashba--spit wire region. As 
it turns out, the phenomena of spin--accumulation and current
conversion are exhibited for either choice of parameters.
\end{thebibliography}
\end{document}